\documentclass{svmult}

\usepackage{amsmath}
\usepackage{amsfonts}
\usepackage{amssymb}
\usepackage{graphicx}
\usepackage{enumerate}
\usepackage{algorithm}
\usepackage{algorithmic, amsmath}
\usepackage{calrsfs}
\DeclareMathAlphabet{\pazocal}{OMS}{zplm}{m}{n}
\usepackage{longtable}
\usepackage{url}
\usepackage{cite}
\usepackage{multicol}        
\usepackage[bottom]{footmisc}
\makeindex  
\begin{document}
\begin{frontmatter}

\title*{The security strength of Blockchain technology : A Survey Report}
\author{Md Arquam\inst{1} \and Ashish Patel  \inst{2} \and Parma Nand\inst{1}}
\authorrunning{Md Arquam,  Ashish Patel , Parma Nand}

\institute{Depatment of Computer Science \& Engineering, \\Sharda University, Greater Noida\\ Uttar Pardesh, India\\
Email: mohd.arquam@sharda.ac.in, parma.nand@sharda.ac.in\\
\and
Depatment of Computer Science \& Engineering, \\Galgotias University, Greater Noida\\ Uttar Pardesh, India\\
Email: ashishp.it.21@nitj.ac.in
}

\maketitle
\vspace*{-.25cm}
\abstract {The advent of blockchain technology by the Nakamoto group in 2008 has created a new trend on how to deal with various security issues and vulnerabilities. Blockchain systems have gained momentum in various spheres of technology deployment in business organizations. This paper presents a critical literature survey on the security strength of blockchains and security issues associated with blockchain technology deployment. Numerous studies have experimented with the various technical features of blockchain systems across various transaction domains. Findings obtained from literature survey and thematic content analysis of the existing research studies indicate that blockchain systems provide unique capabilities that support processes and transactions across various sectors with a high level of integrity, transparency, confidentiality, and privacy. However, some loopholes and limitations associated with deployment and use of blockchains have been highlighted in various studies. The present study cross-examined the security issues of the underlying scientific research evidence. }

\end{frontmatter}
\vspace*{-.8cm}
\section{Introduction}\label{sec:Introduction} 
Blockchain technology is one of the disrupting technologies applied in managing various business processes. It was introduced in 2008 by a group/person called Satoshi Nakamoto to avoid double-spending and solve the challenge of failure to maintain the order of transactions \cite{zahadat2018blockchain}. The breakthrough of the first Bitcoin blockchain has inspired the widespread popularity of blockchain technology across the world. The technology is being used in various applications including e-commerce, supply chain management, data management, education, business, and industrial applications, healthcare management, privacy and security, electronic voting, currency markets, public service, and governance \cite{samaniego2016blockchain}, \cite{arquam2018blockchain} \cite{casino2019systematic}). This has made blockchain technology a subject of scientific research in industrial practitioners, developers, and researchers because of the security and trust features associated with its applications and deployments.  

Despite the growing popularity of blockchain technology, some scholars believe that its security weaknesses have had a lasting impact on the world \cite{choo2015cryptocurrency}, \cite{chen2018econometric}. According to Dorri et al. \cite{dorri2016blockchain} blockchain technology has facilitated the increase in financially motivated cyber-attacks such as the denial of services and ransomware against various organizations and retailers. It has also enabled illicit dark online marketplaces. Today, the adoption and use of blockchain technology are way beyond its initial objective of facilitating the proliferation of decentralized cryptocurrency. Many industries are looking up to its decentralized trustless ledger enriched with a high level of immutability to run business processes such as smart contracts, logistics operations, cybersecurity, banking, and pharmaceutical operations \cite{li2017blockchain}. 

Therefore, the blockchain breakthroughs have caused rigorous changes and disruptions in the ancient business processes especially those that involve transactions and applications that require the verification of trusted third parties and centralized architectures. The blockchain architecture possesses unique characteristics that resonate with unique features such as security, suitability, robustness, and transparency \cite{christidis2016blockchains}. According to Zahadat et. al \cite{zahadat2018blockchain} a blockchain is a distributed database system with a sequence of immutable and committed blocks, which makes them ideal for deployment in the banking sector where customer transactions across various banks can be handled using the same blockchain. This facilitates easy auditing and promotes transparency. Banking corporations are investing in the use of blockchains to decentralize their architectures and reduce transaction costs as they become faster, transparent, and inherently safer than independent architectures. Beyond facilitating cryptocurrency payments, blockchain technology enables coordination of decentralized applications in absence of third parties and intermediaries, a unique facilitating the proliferation of modern internet security systems. 

As blockchain continues to shape business processes and financial transactions, it is imperative to examine its effectiveness in promoting cybersecurity and customer protection against emerging security threats. It is necessary to identify relevant scholarly works and papers on blockchain technology and cybersecurity strength. In this research, we critically review the existing literature on the capacity and security strength of blockchains in supporting applications that necessitate a high level of cyber-security, accountability, integrity, transparency, and privacy. The interplay between cybersecurity and blockchain technology provides insights on the future implementation of applications that utilize its security features. 

\section{Background study}

Blockchains are decentralized and well-distributed time-stamped data structures logically organized to form a distributed peer-to-peer network in which non-trusted members can interact with each other without the authority and verification of a trusted third party \cite{christidis2016blockchains}. Blockchain technology utilizes multiple interconnected mechanisms that provide different feature features to the system infrastructure. The infrastructure has peer-to-peer transactions at the lower end, which demonstrate the consent between two customers conducting business. When one actor signs a transaction, it is widely distributed to the neighbors. The blockchain has entities that connect to it called nodes, which verify its rules. Full nodes arrange transactions into the blocks and determine the validity of the transaction, after which the invalid transactions are discarded when nodes reach a consensus on the valid transactions that should be kept in the chain to ensure that the system has no divergences and corrupt branches \cite{vukolic2015quest}.

According to Mingxiao et al. \cite{mingxiao2017review} different types of blockchains have different consensus mechanisms. These include proof of stake (PoS), proof-of-work (POW), and Byzantine Fault Tolerance (BFT). PoS protocols proportionally segment stake blocks depending on the current wealth of the participants to prevent the richest miner from dominating the network \cite{zheng2018blockchain} . Proof of work protocols splits the blocks proportionally according to the mining power of the participants. The system traces those hashes with specific identical patterns to ensure the verifiability and authentication of transactions. Initially, Ethereum blockchains applied POW protocols but they have also shifted to the proof of stake (PoS) protocols because of their improved scalability and the potential decrease in power consumption \cite{dannen2019introducing}. The third consensus layer called the computing interface enables blockchains to store the system user transactions and compute user balances. 

According to Zahadat et al \cite{zahadat2018blockchain}, advanced computations require the storage of complex states that can be dynamically updated using distributed mechanisms of computing; especially those states that change from one state to another after particular criteria are achieved. The fourth layer of the blockchain called the governance layer expands the capabilities of the blockchain infrastructure to manage physical human interactions across the world. Despite all the defined protocols, blockchains are affected by user inputs from different people who have different intentions hence necessitating the integration of advanced methods to improve the system capabilities. The four layers necessary for the effective functioning of the blockchain handle off-chain social processes. However, the governance layer manages the interactions of various actors to enable them to produce, change, or maintain inputs that buildup the blockchain infrastructure.

Some studies have categorized blockchain networks in various ways according to the network's permissions and management protocols \cite{zheng2018blockchain, wood2014ethereum}. According to Kravchenko \cite{kravchenko2016ok}, private blockchains fall in the permission category in which authorized uses are defined with specific permissions and characteristics over the operations of the network while public blockchains do not involve the use of permissions because everyone is allowed to join as a new node miner or a new user and perform transactions and operations of his choice. Public blockchain features are implemented in most of the existing cryptocurrencies such as Ethereum, Litecoin, and Bitcoin \cite{haferkorn2014seasonality}. 
One of the advantages of using public blockchains is that the network can maintain and sustain itself hence reducing infrastructure management costs. Meanwhile, private blockchain designs have been implemented in performance-oriented systems such as auditing and database management \cite{zheng2018blockchain}. 
Besides, federated blockchains combine some features of private blockchains and public blockchains \cite{zheng2018blockchain}. They share similar privacy protection protocols and scalability levels as private blockchains although they have a different set of nodes called the leading nodes. A combination of the leader nodes is used to verify and validate the transaction processes taking place in the system, instead of using a single entity to perform transaction verification. This feature facilitates the effective implementation of a decentralized system design in which the leader nodes can verify and permit other node miners. Some studies have improved the state of the art of the blockchain network by incorporating more classical features such as information management and ownership, anonymity, and transaction approval time using advanced consensus and security mechanisms of the blockchain. 

\section{Literature survey}
The present literature survey analyzes 20 research studies published between 2015 and 2020 to provide evidence-based insights on the research trends shaping the application of blockchain security features. Various multidisciplinary approaches have been used in modern scientific literature to explain the security strength and effectiveness of blockchain technology. 
\subsection{Datasets: Pros and cons }
We extracted data from papers that meet the criteria for quality assessment. The completeness of data was assessed to determine the accuracy of recorded information. We compile data from both qualitative and quantitative research studies and conduct a meta-analysis. Initial search held with key words such as ‘blockchain’, ‘security’, and ‘bitcoin’ yielded 713 papers. After application of exclusion and inclusion criteria, 131 papers remained but only 20 papers were included for this analysis. Data is extracted from papers that contain empirical evidenced on the use of blockchain technology for security and other applications. The research papers utilized also contain empirical data on the performance of blockchain architectures and distributed ledger technologies. 

Most of the papers are peer reviewed published in a journal or conference proceedings.  Primary studies present relevant quantitative and qualitative data. Each paper is focused on classification of the blockchain security applications, security issues, and technology solutions to address these challenges. Some of the common themes include encrypted data storage, virtual network management, data sharing, bitcon cryptography, cyber security, public key infrastructure, electronic health records, centralized control, data privacy, confidentiality, mining authentication, domain name system, multi-media security, IoT, Big-data, and peer-to-peer sharing. Most of the studies present the various limitations associated with blockchain implementation, propose solutions but the effectiveness of those solution remains theoretical since they lack comprehensive evaluations. 

\subsubsection{Key findings from the literature survey}

\begin{longtable}{| p{3.0cm}|| p{3.0cm} || p{6.0cm} |} 
\hline
\textbf{Authors } & \textbf{Types of Security Applications}   & \textbf{ Quantitative/Qualitative  Data Reported} \\ \hline
Pinno et al (2017) & IoT & Presents a design of a “ControlChain system”. This is a blockchain based solution for IoT device access control using the bitcon principles. Found that multiple blockchains can be utilized to manage various   aspects of the IoT controls. \\ \hline
Banerjee et al & IoT   & The paper presents the increasing importance of implementing blockchain technology in healthcare, battlefield, and IoT homes. Postulates how IoT can install a robust-secure firmware using blockchains for firmware updates. 
\\ \hline
Casino, Dasaklis, \& Patsakis (2019) & Classiﬁcation and main characteristics of blockchain networks and Taxonomy of blockchain-based applications  
 & E-commerce, supply chain management, data management, education, business, and industrial applications, healthcare management, privacy and security, electronic voting, currency markets, public service, and governance  \\ \hline 
 Gu et al (2018) &  Android  Malware &
Explains the working of blockchain in detecting android malware. Utilizes a set of blockchains where the intrusion detection system has specified N members to detect hashed malware on Android devices. \\ \hline
Bhowmik and Feng (2017) &  Multimedia security & The paper presents the application of blockchain technology in multimedia watermarking. The authors present a blockchain framework for multimedia watermarking to address the underlying issues. Water mark information contains an image hash preserving the original media content and the cryptographic hash containing transaction histories (transaction log). After extraction, the watermark extract is passed to a distributed ledger to retrieve the transaction log and the latter segment is applied to identify the tampered/edited regions. \\ \hline
Shi, He, Li, Kumar, Khan \& Choo, (2020) & Electronic health record (EHR) & Presents a systematic analysis of the blockchain models applied in Electronic health record systems (EHRS). Emphasis is placed on data privacy and security aspects; background knowledge relating to blockchain and EHR systems; applications of blockchain in EHR systems; Research challenges and opportunities. \\ \hline
Laskowski and Kim (2016) & Big Data, IoT &  The paper presents the blockchain architecture; how it can be used in provenance tracking. The study applies blockchain ontologies on knowledge traceability, knowledge provenance, and food provenance. Such ontologies can be applied in designing a blockchain system. The authors analyze the contribution of traceability ontology and how it can be applied smart contracts to enforce traceability constraints and execute a provenance trace in Ethereum blockchain platforms. \\ \hline
Xun et al. (2017)  & IoT & The study investigates various security attacks; and how they affect physical processes, and then propose effective blockchain-based detection mechanisms. We depict two different types of command disaggregation attack modes: disaggregated sub-commands are allocated to wrong actuators and the command sequence is disordered. Three models to implement command disaggregation attack modes. The designed framework use  the relationships in the two-tier command sequences, including sub-commands from the input of actuators  and commands from the output of central controller to detect attacks before disruptions occur. \\ \hline 
Zolanvari, Erbad, Jain, \& Samaka, (2018) & Centralized controllers & Presents a survey of the various technology solutions designed using blockchain models to provide security services such as include confidentiality, authentication, privacy and access control list, integrity assurance, data and resource provenance. The paper examines the use of blockchain security services for modern business applications. Highlights challenges associated with using blockchain-based security services Highlight the state of the art techniques utilized to provide these services, their challenges, explain the modern use of blockchain technology can resolve these challenges. Compares different blockchain-based techniques and frameworks providing security services.   \\ \hline 
Arquam, Singh \& Sharma (2018)  & Security in Online Social Network & Proposes  a blockchain based model for sharing the information securely at the peer level.  In the proposed model, a chain is created by combining blocks of information generated by nodes. Each node in the network propagates the information based on its credibility against its peer nodes. The credibility of a node varies according to their respective information. Trust is calculated between sender and receiver using either of two ways, Local trust and Global trust.   \\ \hline
Alvarenga, Rebello, \& Duarte (2018) & Virtual Network Management & Presents the blockchain enabled security of virtual networks and management of data centers. A consensus blockchain technology solution is proposed to apply a PBFT consensus mechanism. The experimental write speed for PBFT system are 10–20 times higher than the normal write speeds obtained using bitcoin and ethereum platforms.   \\ \hline 
Khan \& Salah, (2018) & IoT & The study examines IoT security challenges; how blockchain technology solutions can meet such challenges and reduce security threats. Ethereum is utilized as target platform for smart contracts.   \\ \hline
Yli-Huumo et al., (2016) & IoT & Discusses the security strength of blockchain technology in enhancing IoT security. Presents the security benefits for supply chain systems. Explore various technology issues and the future direction of blockchain technology. Many proposed technology solutions lack comprehensive evaluation for effective deployment. Blockchain technology presents issues in terms of latency, scalability, and throughput.    \\ \hline
   Devetsikiotis and Christidis (2016) & IoT/Smart Contracts &
 The paper examines the operation of blockchain mechanisms focusing on smart contract-script for multi-step process automation. The study also extends to IoT domains and examines the possibilities for integration of blockchain technology in IoT for various purposes such as cryptographic automation, and sharing of resources and services between devices.   \\ \hline
   Suhaliana et al (2018) & N/A & The paper presents the various technology applications of where blockchain can be integrated including security, media, and finance. Discusses the security issues and proposes theoretical security solutions to address the highlighted issues. The paper gives a future direction for research in the security aspects of blockchain.        \\ \hline        
Xu et al, (2017) & Web Applications & Proposes the use of a Distributed Ledger Based Access Control (DL-BAC) for navigation and management of web applications. Distributed ledger is constructed in the generic blockchain architecture similar that of a bitcoin.    \\ \hline          
 Cai, C., Yuan, X., \& Wang, C. (2017) & Encrypted Data Storage \& Searching & Blockchain based distribution of hashed search indices to allow for keyword searching of encrypted data. Integrity maintained by obtaining value deposit from a joining user and if they act maliciously, this deposit is shared to the rest of the nodes.          \\ \hline
Moinet, Darties,  \& Baril (2017) & Public Key Infrastructure & Proposes a blockchain model to facilitate the management of public key infrastructure. Node mining is enhanced through data payloads that are labelled revoke, approve, ban, blame, and renew and auth. These payloads build trust across nodes.            \\ \hline
            
\caption{Important research in Blockchain Technology}
\label{table:1}
\vspace{-.4cm}
\end{longtable}

Across various research studies, blockchain systems are decentralized and truly distributed systems developed to provide solutions to various related problems. Some of the practical security solutions provided in the primary studies demonstrate innovative approaches to solving contemporary challenges associated with user authentication, immutability, and data security \cite{shi2020applications}. These technology solutions depend on the changes in the network infrastructure introduced by blockchain technology compared to the traditional use of one centralized server. Because of the costs and labor associated with moving or changing a working system, some of the practical security concepts cannot be easily experimented with within a particular limit of time to determine the security strength of blockchain applications. 

Robinson \cite{robinson2018requirements} experimented with various blockchain consensus mechanisms using the IoT chain design. They used an Ethereum platform for experimental and developmental analysis. The authors deduced that the most ready-to-deploy and practical security solutions are those experimented on Ethereum platforms.  
Blockchains provide unique capabilities for business transactions to proceed with minimal cybersecurity issues. According to Casino et al. \cite{casino2019systematic}, blockchains support the inherent organizational strategies to secure their information/data, communications, and networks. They utilize hashing and encryption algorithms to store large size of immutable records and provide security/protection against intrusion. 

According to Taylor et al \cite{taylor2020systematic}, most of the security measures deployed in various organizations depend on one trusted source to store the encrypted data and verify the information. Consequently, such systems are susceptible to intruders and malicious attackers who target them cause a lot of damages through data extortion, injection of malicious information, blackmail, information theft, and denial of service attacks. Fortunately, blockchain technologies provide decentralized security solutions to such system weaknesses. Its solutions do not need the trust or authority of any member of the network or transaction group. Each member of the network or each node connected to the system has a full copy of the historic records hence requires the consensus of the majority nodes to add more data into the chain of previous records. 

According to Taylor et al \cite{taylor2020systematic}, Consensus-based verification is achieved in various ways although there is a bottom line. Group members with accessibility to the same information can secure their group better than compared to the group composed of a host of members and one leader who depend on their leader for information access in presence of malware that comes into the system as leaders or group members.    
Some scholars have discussed how the capabilities of blockchain technology can be leveraged to improve cybersecurity in various applications such as data sharing, data storage, internet of things and, World Wide Web navigation, private user data management, and network security. According to Pinno et al \cite{pinno2017controlchain}, many IoT applications use private blockchain structures to manage controlled access for network nodes, prevent malicious intrusion, and securely track data management. Besides, Banerjee et al \cite{banerjee2018blockchain} found that blockchain systems can facilitate the propagation of updates from one node to another hence enhancing the security of firmware deployment. This simplifies the authentication and identification of IoT devices and facilitates seamless data transfers across the system. Gu et al \cite{gu2018consortium} have also experimented and confirmed the same findings on the effectiveness of blockchain technology in detecting malicious behavior and securing historic IoT sessions and connections. In these research studies, the authors suggest conventional architecture in which blockchain protocols exist between the transport and application layers hence utilize/treat rewards as units of the voting power.

One of the potential capabilities of blockchains is its support for integrity verification. The effectiveness and capabilities of blockchain technology have been studied in various studies \cite{zikratov2017ensuring}, \cite{bhowmik2017multimedia}. The study conducted by Bhowmik et al \cite{bhowmik2017multimedia} on the integrity verification capabilities of blockchains found that applications that involve blockchain integrity verification such as intellectual property management, insurance, counterfeit, and provenance systems can store transactions and information related to the lifetime and creation of services or products. According to DeLaRosa et al \cite{de2017survey}, blockchain applications that can accept integrity verification are those that are based on IP protection. Technology solutions such as Mediachain and Ascribe use blockchain technology to connect digital content to their innovator. For ascribe, bitcoin blockchains are used to transfer the loan and the ownership of loaned digital assets. Meanwhile, Mediachain uses blockchain technology to store metadata for easy media querying and recovery.

Besides, some of the modern monetization techniques such as monegraphs allow revenue sharing across the media value chain for image reals, video clips, online broadcasts, and more sponsored content when all the verifications are done by the blockchain technology \cite{monegraph}. Some blockchain solutions such as Factom provide validation and storage of digital assets \cite{snow2015factom}. Meanwhile, Kodakcoin blockchains are facilitating payments for the acquisition of image rights and photo licenses via KodakOne blockchain –enabled platform that stores all the content for its registered photographers \cite{corbet2020kodakcoin}. In information service, Silent Notary is an example of a blockchain-enabled service for confirmation of the existence of events that are digitally recorded and content stored in digital formats such as on e-mail, video file, and image file \cite{casino2019systematic}. 

Despite the proliferation of these blockchain-enabled solutions, Herbaut et al \cite{herbaut2017model} insist that there should be a user-centered technology shaping the inherent ecosystem for content delivery. The study conducted by Laskowski et al \cite{kim2018toward} explains the philosophical perspectives of storing and interpreting data automatically to preserve data integrity and provenance. Authors emphasize that such Smart contracts have related ontologies of data storage and interpretation that can be adopted to use blockchain technologies. Today, technology solutions such as BlockVerify and Everledger use a combination of smart contracts and blockchains to avoid counterfeits and bank fraud for insurances and banks as well as promoting supply chain transparency \cite{lomas2015everledger}. In the same accord,   Xun et al. \cite{xun2017command} emphasize data integrity in their study which implemented blockchain protocols in their prototype system aimed at demonstrating a blockchain technology framework for promoting data integrity in service delivery. Besides, Jaag et al. \cite{jaag2017blockchain} also demonstrate the possibilities of using blockchains in device management, identity services, and supply chain management in business organizations.  

 Blockchain technology has gained more popularity for the insurance sector in various data integrity services such as asset transfers, premium payments, claims processing, customer onboarding, underwriting, sales management, and reinsurance \cite{KPMG}. The recent launch of the B3i-blockchain initiative for the insurance sector by European-based insurers exemplifies how the increasing deployment of blockchain technology in developing standards and processes for the insurance sector to increase efficiency in service delivery \cite{cognizant2017potential}. Blockchain smart contracts lead to process automation in the insurance service delivery thus increasing processing speed, efficiency, and reduced costs. The adoption of blockchain technology has been largely witnessed in the health insurance sector where the need for high-level data security for wellness and medical information is highly cherished.

Concerning data sharing and storage, blockchain ledgers especially private and public blockchains are being used to eliminate circumstances such as "single failed source" within the data storage ecosystem by securing data integrity \cite{xu2017dl}. According to Casino et al \cite{casino2019systematic}, Blockchains protects the integrity of data stored in the cloud by ensuring its resistance against unauthorized changes. They also facilitate the use of hash lists that enable secure data searching and verified data exchange as data structures remain stored and maintained securely. The capability of blockchain technology to create a decentralized network improves data sharing security and data storage especially in those networks that utilize client-based encryptions where the owners of information have full access control over their information \cite{shi2020applications}. Blockchain technologies are also being in securing the utilization and navigation of the World Wide Web by improving the verification and validity of wireless internet access points that are linked to the distributed decentralized network. The technology utilizes hash algorithms programmed to monitor and store access control information local ledgers \cite{salman2018security}. In World Wide Web navigation, blockchains identify the correct web pages and help the user to navigate through correct domain server records, safely communicate with other web users through encrypted/secure methods, and safely utilize web applications \cite{casino2019systematic}. This technology solution employs consortium blockchains where the consensus processes are controlled by predetermined sets of network nodes connected on the distributed network.

The effectiveness of blockchain technology on promoting network security has also been studied in software-defined networks \cite{alvarenga2018securing}.  In these research studies, SDN controllers inbuilt with blockchain technology use cluster structures to facilitate peer to peer communications. Private and public blockchains are used to facilitate node communications in the SDN controllers and the distributed network hence the network system is enriched with capabilities to handle the underlying network security issues. Containers are used for critical data authentication for the decentralized data stores. Blockchain solutions are also enriched with capabilities to manage network security without cryptocurrency tokens. Several research studies agree that token incentivization for all node miners is a robust technique for achieving the longest chain consensus \cite{khan2018iot}. Currency Tokens allow receiver nodes to have more voting powers yet the voting power depends on the contribution of nodes to mining. According to Patrício \& Ferreira \cite{patricio2020blockchain}, it is possible for each IoT device to automatically charge a currency token to other devices for influencing firmware upgrades. To address this challenge, Taylor, et al \cite{taylor2020systematic} suggests the use of many blockchain layers to check the authentication and trust verification of transactions between various hierarchical layers.

\subsection{Security issues faced in the implementation of blockchain technologies}

Despite the aforementioned security capabilities and application of blockchain technologies in the context of data storage, cybersecurity, and private data management, several studies have highlighted several security weaknesses and limitations of using blockchain technologies \cite{yli2016current}, \cite{salman2018security}; 
According to Yli-Huumo et al \cite{yli2016current}, maintaining user confidentiality and privacy remains a key challenge in blockchain-based systems because the information being accessed is stored as a public ledger. This requires advanced encryption and anonymization mechanisms to protect the privacy and confidentiality of the stored information on the public ledger. Unfortunately, such mechanisms depend on the system implementation contexts and may necessitate additional requirements for the networks to operate effectively. According to Devetsikiotis and Christidis \cite{christidis2016blockchains}, secure protocols should be used to perform network file sharing and prevent unverified information disclosure. Other mechanisms to promote data confidentiality suggested by scientific research studies include direct use of a peer-to-peer file system that is content-addressed to facilitate advanced verifications \cite{huckle2016internet}.

According to Suhaliana et al., (2018), blockchains have weaknesses in maintaining transaction privacy. Many system users and businesses expressed concerns about the traceability of Smart contract operations and transactions propagating across the decentralized distributed networks. In the same agreement, Kosba et al. \cite{kosba2016hawk} noted that the use of pseudonyms in blockchain systems as a security measure cannot sufficiently guarantee transaction privacy. As an alternative, Meiklejohn et al \cite{meiklejohn2013fistful},  suggest the use of deanonymisation techniques that examine the transactional structures of the flowing cryptocurrencies. According to Goldfeder et al \cite{goldfeder2018cookie}, the transactions made with bitcoin blockchains can relinquish more sensitive information; hence compromising the privacy and confidentiality of the transaction. Besides, Smart contracts usually contain errors because they resemble the implementation programs hence causing heavy losses to the customers. For example, the Parity wallet bug that happened in 2017 made users lose about 280 million pounds while the DAO attack generated a loss of about 47 million pounds. Recently, Pearson technologies discovered thousands of vulnerable Smart contracts \cite{siegel2016understanding} (Siegel, 2016; Pearson, 2018). The complex nature of smart contracts makes it difficult for technicians to understand their operations. According to Bartoletti et al \cite{bartoletti2020dissecting}, smart contracts are different from conventional programming environments, which make it possible for them to hide illicit transaction behaviors such as Ponzi scams. Ponzi scams have been criticized for disguising as serious investment opportunities promising high business returns yet they compromise private hashes and keys and fail to return funds to the subscribers after failing to meet their business goals \cite{bartoletti2020dissecting}. According to Dannen \cite{dannen2019introducing}, the most promising approach to counter most of the smart contract abuses and vulnerabilities is the one that limits the expressiveness of the intrinsic programming language. Some security solutions utilize Smart contract checkers to implement a framework for verifying the fairness and correctness of the smart contracts hence they can trace underlying threats and vulnerabilities \cite{nikolic2018finding}.

The deployment of blockchain technology on distributed networks is also limited by throughput issues. Koteska, Karafiloski, \& Mishev \cite{koteska2017blockchain} explain the various limitations and challenges associated with blockchain technology features in which he highlights several throughput issues. The current bitcoin networks process a maximum of three to twenty transactions per second. Such a limitation cannot allow the use of bitcoins in systems where the number of user requests exceeds the system's throughput. Instead, users can resort to alternative networks such as the Twitter network which has a throughput of 5000 transactions per second, or VISA transaction networks that can process more than 5000 transactions per second. Blockchain network operations are also limited by latency issues. Users expect their requests to be processed immediately over the internet yet there is a universal acceptance obstacle. The amount of time required to complete a transaction within the bitcoin block is approximately ten minutes. Such delays are attributed to security checks and verifications associated with the transaction (Beck et al. 2016). Larger transactions can spend almost one hour before they are processed, which subjects them to double spending attacks yet other possible alternatives such as Visa transactions can take a maximum of seconds \cite{yli2016current}. The bandwidth and the size of the blockchain have also been questioned in some research studies. According to Yli et al, the current size of the bitcoin INS 1Megabites and it takes ten minutes to create a new block. This implies that one block can handle an average of 500 transactions in 24hours. Therefore, the blockchain cannot handle more transactions than its size hence it cannot be deployed in applications whose transactions exceed its capacity.

Blockchain systems have also been criticized for having scalability issues. Appropriate and secure implementation of blockchains to provide security requires blockchains with multiple full nodes such that the technology implementation may not result in a less decentralized distributed system \cite{cai2017hardening}. The scalability limitations of the blockchain systems are associated with the data size on the blockchain, the latency of data transmission, and the transaction processing rate. Bitcoin latency between transaction confirmation and transaction submission is approximately one hour while Ethereum takes approximately three hours to confirm a transaction \cite{xu2017dl}. Blockchains systems are also criticized for increased cost issues associated with system decentralization and deployment \cite{koteska2017blockchain}. The users of these systems pay for computational power and transaction charges. Unfortunately, the systems do not alert users of the associated charges since prices remain hidden. Another issue affecting/limiting the implementation or deployment of blockchain technologies is data malleability. According to Beck et al \cite{beck2016blockchain}, blockchain data security signatures do not guarantee the bitcoin ownership for a given bitcoin transfer with a transaction. This implies that a malicious attacker can alter and rebroadcast the transaction and cause difficulties in the transaction confirmation \cite{yli2016current}. Authentication challenges have also been highlighted in various studies. Transaction authentication remains a key security concern regarding the implementation of blockchains \cite{koteska2017blockchain}. Customers/users are not secure from malicious attackers who compromise customer private keys and steal the information or alter the transactions. Privacy issues have been highlighted in various scholarly sources. Blockchain systems allow users to create multiple addresses and cluster these user addresses \cite{herrera2016privacy}. The purpose of clustering user addresses is to identify the movements and behaviors of the same system users by identifying all the addresses created in the system by the same user \cite{reid2013analysis}. Koshy et al \cite{koshy2014analysis} examined the effectiveness of bitcoin systems in handling multiple addresses and found that it is possible to map some of the bitcoin addresses to the IP addresses by monitoring and examining the associated transaction traffic.

The susceptibility of blockchain systems to double-spending attacks limits its holistic implementation in various applications. According to Casino, Dasaklis, \& Patsakis \cite{casino2019systematic}, a double-spending attack in bitcoin systems happens when an intruder hoards his bitcoin as he receives services that are yet to be re-spent. For example, when an intruder credits a particular account, he/she receives the goods or services from the owner of the account. Thereafter, he/she will revert the transaction crediting the account to recognize the associated ledger. However, some studies have shown that bitcoin systems can control or prevent the occurrence of double-spending attacks using rigorous models that reframe the honest users and the attacker to appear as competitors performing random movements in one direction each with probabilistic steps ahead of the other. Nevertheless, Garay et al \cite{garay2015bitcoin} emphasize that the decentralized nature of blockchain systems makes it easier for an attacker to disagree with honest node miners.

The increased dependence of blockchain systems on distributed public ledgers has raised some security concerns. According to Xu et al \cite{xu2017dl}, Public distributed ledgers have speculative issues associated with the compromise between decentralization and network dimensions. The probability of intrusion in bitcoin blockchain ledgers is approximately 51\% because a single node miner can easily gain full control of the biggest size of the network resources thus holding other users at ransom. Several studies have confirmed various security compromises and currency scams associated with the implementation of bitcoin blockchains \cite{lim2014analysis}. These include account hacking using viruses and Trojan horses from online adverts and distributed denial of service attacks. This leaves many bitcoin miners susceptible to scams from malicious attackers. Besides, the proof of work protocols inherent in bitcoin chains has been criticized for resource wastage \cite{yli2016current}. The work done by the node miner determines his/her possibility for mining a block. Besides, small nodes are at higher risk of being attacked by malicious intruders. It is also difficult to combine various segments of blockchains for versioning and administrative purposes.

Some issues have also been noted in the blockchain quantum computing capabilities. The blockchain core utilizes public key encryptions and hashes as cryptographic primitives used to verify and authenticate transactions. Initial blockchain infrastructures utilized quantum computing; but several technology breakthroughs led to the introduction of SHA-256, which is a hash algorithm used to perform quantum computing. To crack this technology using the Grover algorithm, an intruder needs around 2128 operations. The resilience of SHA-256 against quantum attacks does not apply in those blockchains that utilize public key encryption algorithms. According to Casino, Dasaklis, \& Patsakis \cite{casino2019systematic}, an intruder can break the ECDSA algorithm using a big quantum computer hence increasing the vulnerability of blockchains. As quantum computing resilience becomes a major issue, there is a need to standardize blockchain systems such that they can perform post-quantum cryptographic analysis. Recent studies such as Kiktenko et al \cite{kiktenko2018quantum} have developed quantum secure blockchain systems that utilize public key distribution across the optical fiber network for secure authentication. Meanwhile, Visser and Rajan \cite{rajan2019quantum} proposed blockchain encoding into a temporary state of non-existent photons so that the entire blockchain system can be perceived as a quantum interconnected machine.

\section{Conclusions}

This study has surveyed a wide range of multidisciplinary studies on the security strengths and effectiveness of blockchain deployments in various applications. Unquestionably, blockchains have proven their capability of improving and transforming the daily business processes and transactions in various businesses and organizations. The technology has gained widespread adoption in many countries across the world. Its benefits are far beyond the security issues associated with the use of centralized database systems. Despite the increased deployment and use of blockchains, several studies have has highlighted various security issues and other limitations that need to be properly addressed if blockchain technology is to thrive in the drastically changing and competitive tech market. By addressing the above issues, blockchains will increasingly become more efficient and scalable for both the users and the organizations deploying them. Currently, blockchains possess numerous mechanisms that make them the best choice for use in various applications in several industries and business sectors. The future presents more bright opportunities for blockchains to penetrate various business domains and industries although the compromise and trade-off lie between the use of blockchains and conventional centralized databases systems. Some of the highlighted security issues also cut across traditional database systems, which makes it difficult for some organizations to shift to the implementation and use of blockchain systems. Besides, each application domain requires unique security characteristics and system behaviors to provide security capabilities tailored to the needs of the application, which does not guarantee system interoperability. There is a need for more innovations in the field of blockchain research and development to reduce energy consumption and increase transaction speed. Protocols should be globally accepted so that the system can easily verify the identities of various individuals at the same time.

What makes blockchain systems outstanding is the ontology that data/information is not owned by one company or individual. Each node minder has access to his/her entire transaction records, which creates a difference between blockchain systems and the traditional database management systems used by various institutions, government agencies, and corporations to store and maintain private records. In traditional database systems, the business organization has full control and access to the records of the clients, customers, and internal operations. In contrast, blockchain technologies record transactions in cryptal formats hence each user shares the same transaction lists that are periodically updated to ensure the transaction ledger is kept current. For every new transaction taking place in the system, the transaction ledgers are reconciled, which minimizes the risk of transaction loss. This implies that that blockchains can maintain the integrity of transaction history because of their distributed P2P network. An attacker or an intruder needs to first change most of the hash values and data blocks within the system to alter the transaction history, which necessitates him/her to compromise many computers of system users across the world.  

Blockchain systems support the existing technical efforts to protect data integrity, secure communications, and networks. Using hashing and encryption, blockchains can store immutable transaction records. The weakness of the traditional database system lies in its features such as utilizing only one trusted authority to store encrypted data and verify information, which subjects the users to the risk of malicious attacks. Many intruders and illicit system users can combine their efforts against the single authority to inject malicious information, commit denial of service attacks, and compromise data integrity through blackmail or theft. However, blockchains do not require the use of a trusted individual or authority because they are decentralized to the user. There no need for the trust since each member of the network (node) has a full copy of his transaction history available. The system only needs to first achieve the majority nodes' consensus to allow more addition of data to the blockchain of the already existing information. This capability has increased the popularity of blockchain and facilitated its deployment in various security-oriented applications such as data sharing and storage, private user data management, network security, and cyber security in the internet of things. Its immutable decentralized database benefits both businesses and consumers. As benefits benefit from minimized costs associated with information storage and maintenance, customers will maintain ownership of their transaction records. The self-monitoring and automatic activity recording promote record availability, integrity, confidentiality, and transparency. Customers using blockchain-enabled systems have less fear over the theft of their data from malicious attacks. Besides, blockchains are providing solutions that address the increasing cyber-attacks by ensuring data/information availability through redundancy, confidentiality through cryptography, and integrity through self-validation.

\vspace{-.8cm}
\bibliographystyle{unsrt}
\bibliography{references}

\begin{thebibliography}{10}

\bibitem{zahadat2018blockchain}
N~Zahadat and W~Partridge.
\newblock Blockchain: A critical component to ensuring data security.
\newblock {\em Journal of Forensic Sciences \& Criminal Investigation}, 10(1),
  2018.

\bibitem{samaniego2016blockchain}
Mayra Samaniego, Uurtsaikh Jamsrandorj, and Ralph Deters.
\newblock Blockchain as a service for iot.
\newblock In {\em 2016 IEEE international conference on internet of things
  (iThings) and IEEE green computing and communications (GreenCom) and IEEE
  cyber, physical and social computing (CPSCom) and IEEE smart data
  (SmartData)}, pages 433--436. IEEE, 2016.

\bibitem{arquam2018blockchain}
Md~Arquam, Anurag Singh, and Rajesh Sharma.
\newblock A blockchain based secure and trusted framework for information
  propagation on online social networks.
\newblock {\em arXiv preprint arXiv:1812.10508}, 2018.

\bibitem{casino2019systematic}
Fran Casino, Thomas~K Dasaklis, and Constantinos Patsakis.
\newblock A systematic literature review of blockchain-based applications:
  current status, classification and open issues.
\newblock {\em Telematics and informatics}, 36:55--81, 2019.

\bibitem{choo2015cryptocurrency}
Kim-Kwang~Raymond Choo.
\newblock Cryptocurrency and virtual currency: Corruption and money
  laundering/terrorism financing risks?
\newblock In {\em Handbook of digital currency}, pages 283--307. Elsevier,
  2015.

\bibitem{chen2018econometric}
Shi Chen, Cathy Yi-Hsuan Chen, Wolfgang~Karl H{\"a}rdle, Teik~Ming Lee, and
  Bobby Ong.
\newblock Econometric analysis of a cryptocurrency index for portfolio
  investment.
\newblock In {\em Handbook of Blockchain, Digital Finance, and Inclusion,
  Volume 1}, pages 175--206. Elsevier, 2018.

\bibitem{dorri2016blockchain}
Ali Dorri, Salil~S Kanhere, and Raja Jurdak.
\newblock Blockchain in internet of things: challenges and solutions.
\newblock {\em arXiv preprint arXiv:1608.05187}, 2016.

\bibitem{li2017blockchain}
Cheng Li and Liang-Jie Zhang.
\newblock A blockchain based new secure multi-layer network model for internet
  of things.
\newblock In {\em 2017 IEEE International congress on Internet of Things
  (ICIOT)}, pages 33--41. IEEE, 2017.

\bibitem{christidis2016blockchains}
Konstantinos Christidis and Michael Devetsikiotis.
\newblock Blockchains and smart contracts for the internet of things.
\newblock {\em Ieee Access}, 4:2292--2303, 2016.

\bibitem{vukolic2015quest}
Marko Vukoli{\'c}.
\newblock The quest for scalable blockchain fabric: Proof-of-work vs. bft
  replication.
\newblock In {\em International workshop on open problems in network security},
  pages 112--125. Springer, 2015.

\bibitem{mingxiao2017review}
Du~Mingxiao, Ma~Xiaofeng, Zhang Zhe, Wang Xiangwei, and Chen Qijun.
\newblock A review on consensus algorithm of blockchain.
\newblock In {\em 2017 IEEE international conference on systems, man, and
  cybernetics (SMC)}, pages 2567--2572. IEEE, 2017.

\bibitem{zheng2018blockchain}
Zibin Zheng, Shaoan Xie, Hong-Ning Dai, Xiangping Chen, and Huaimin Wang.
\newblock Blockchain challenges and opportunities: A survey.
\newblock {\em International Journal of Web and Grid Services}, 14(4):352--375,
  2018.

\bibitem{dannen2019introducing}
Chris Dannen.
\newblock Introducing ethereum and solidity, 2019.

\bibitem{wood2014ethereum}
Gavin Wood et~al.
\newblock Ethereum: A secure decentralised generalised transaction ledger.
\newblock {\em Ethereum project yellow paper}, 151(2014):1--32, 2014.

\bibitem{kravchenko2016ok}
Pavel Kravchenko.
\newblock Ok, i need a blockchain, but which one.
\newblock {\em Medium Corporation}, 2016.

\bibitem{haferkorn2014seasonality}
Martin Haferkorn and Josu{\'e} Manuel~Quintana Diaz.
\newblock Seasonality and interconnectivity within cryptocurrencies-an analysis
  on the basis of bitcoin, litecoin and namecoin.
\newblock In {\em International Workshop on Enterprise Applications and
  Services in the Finance Industry}, pages 106--120. Springer, 2014.

\bibitem{shi2020applications}
Shuyun Shi, Debiao He, Li~Li, Neeraj Kumar, Muhammad~Khurram Khan, and
  Kim-Kwang~Raymond Choo.
\newblock Applications of blockchain in ensuring the security and privacy of
  electronic health record systems: A survey.
\newblock {\em Computers \& Security}, page 101966, 2020.

\bibitem{robinson2018requirements}
Peter Robinson.
\newblock Requirements for ethereum private sidechains.
\newblock {\em arXiv preprint arXiv:1806.09834}, 2018.

\bibitem{taylor2020systematic}
Paul~J Taylor, Tooska Dargahi, Ali Dehghantanha, Reza~M Parizi, and
  Kim-Kwang~Raymond Choo.
\newblock A systematic literature review of blockchain cyber security.
\newblock {\em Digital Communications and Networks}, 6(2):147--156, 2020.

\bibitem{pinno2017controlchain}
Otto Julio~Ahlert Pinno, Andre Ricardo~Abed Gregio, and Luis~CE De~Bona.
\newblock Controlchain: Blockchain as a central enabler for access control
  authorizations in the iot.
\newblock In {\em GLOBECOM 2017-2017 IEEE Global Communications Conference},
  pages 1--6. IEEE, 2017.

\bibitem{banerjee2018blockchain}
Mandrita Banerjee, Junghee Lee, and Kim-Kwang~Raymond Choo.
\newblock A blockchain future for internet of things security: a position
  paper.
\newblock {\em Digital Communications and Networks}, 4(3):149--160, 2018.

\bibitem{gu2018consortium}
Jingjing Gu, Binglin Sun, Xiaojiang Du, Jun Wang, Yi~Zhuang, and Ziwang Wang.
\newblock Consortium blockchain-based malware detection in mobile devices.
\newblock {\em IEEE Access}, 6:12118--12128, 2018.

\bibitem{zikratov2017ensuring}
Igor Zikratov, Alexander Kuzmin, Vladislav Akimenko, Viktor Niculichev, and
  Lucas Yalansky.
\newblock Ensuring data integrity using blockchain technology.
\newblock In {\em 2017 20th Conference of Open Innovations Association
  (FRUCT)}, pages 534--539. IEEE, 2017.

\bibitem{bhowmik2017multimedia}
Deepayan Bhowmik and Tian Feng.
\newblock The multimedia blockchain: A distributed and tamper-proof media
  transaction framework.
\newblock In {\em 2017 22nd International Conference on Digital Signal
  Processing (DSP)}, pages 1--5. IEEE, 2017.

\bibitem{de2017survey}
Josep~Lluis De~La~Rosa, Victor Torres-Padrosa, Andr{\'e}s El-Fakdi, Denisa
  Gibovic, O~Horny{\'a}k, Lutz Maicher, and Francesc Miralles.
\newblock A survey of blockchain technologies for open innovation.
\newblock In {\em Proceedings of the 4th Annual World Open Innovation
  Conference}, pages 14--15, 2017.

\bibitem{monegraph}
Monegraph uses bitcoin tech so internet artists can establish “original”
  copies of their work.
\newblock In {\em https://techcrunch.com/2014/05/09/monegraph/}. Monegraph,
  2014.

\bibitem{snow2015factom}
Paul Snow, Brian Deery, Peter Kirby, and David Johnston.
\newblock Factom ledger by consensus, 2015.

\bibitem{corbet2020kodakcoin}
Shaen Corbet, Charles Larkin, Brian Lucey, and Larisa Yarovaya.
\newblock Kodakcoin: a blockchain revolution or exploiting a potential
  cryptocurrency bubble?
\newblock {\em Applied Economics Letters}, 27(7):518--524, 2020.

\bibitem{herbaut2017model}
Nicolas Herbaut and Nicolas Negru.
\newblock A model for collaborative blockchain-based video delivery relying on
  advanced network services chains.
\newblock {\em IEEE Communications Magazine}, 55(9):70--76, 2017.

\bibitem{kim2018toward}
Henry~M Kim and Marek Laskowski.
\newblock Toward an ontology-driven blockchain design for supply-chain
  provenance.
\newblock {\em Intelligent Systems in Accounting, Finance and Management},
  25(1):18--27, 2018.

\bibitem{lomas2015everledger}
Natasha Lomas.
\newblock Everledger is using blockchain to combat fraud, starting with
  diamonds.
\newblock {\em Tech Crunch}, 29, 2015.

\bibitem{xun2017command}
Peng Xun, Pei-Dong Zhu, Yi-Fan Hu, Peng-Shuai Cui, and Yan Zhang.
\newblock Command disaggregation attack and mitigation in industrial internet
  of things.
\newblock {\em Sensors}, 17(10):2408, 2017.

\bibitem{jaag2017blockchain}
Christian Jaag and Christian Bach.
\newblock Blockchain technology and cryptocurrencies: Opportunities for postal
  financial services.
\newblock In {\em The changing postal and delivery sector}, pages 205--221.
  Springer, 2017.

\bibitem{KPMG}
KPMG International.
\newblock Blockchain accelerates insurance transformation.
\newblock
  https://assets.kpmg/content/dam/kpmg/xx/pdf/2017/01/blockchain-accelerates-insurance-transformation-fs.pdf,
  2017.

\bibitem{cognizant2017potential}
Blockchain Cognizant.
\newblock A potential game-changer for life insurance, 2017.

\bibitem{xu2017dl}
Lei Xu, Lin Chen, Nolan Shah, Zhimin Gao, Yang Lu, and Weidong Shi.
\newblock Dl-bac: Distributed ledger based access control for web applications.
\newblock In {\em Proceedings of the 26th International Conference on World
  Wide Web Companion}, pages 1445--1450, 2017.

\bibitem{salman2018security}
Tara Salman, Maede Zolanvari, Aiman Erbad, Raj Jain, and Mohammed Samaka.
\newblock Security services using blockchains: A state of the art survey.
\newblock {\em IEEE Communications Surveys \& Tutorials}, 21(1):858--880, 2018.

\bibitem{alvarenga2018securing}
Igor~D Alvarenga, Gabriel~AF Rebello, and Otto Carlos~MB Duarte.
\newblock Securing configuration management and migration of virtual network
  functions using blockchain.
\newblock In {\em NOMS 2018-2018 IEEE/IFIP Network Operations and Management
  Symposium}, pages 1--9. IEEE, 2018.

\bibitem{khan2018iot}
Minhaj~Ahmad Khan and Khaled Salah.
\newblock Iot security: Review, blockchain solutions, and open challenges.
\newblock {\em Future Generation Computer Systems}, 82:395--411, 2018.

\bibitem{patricio2020blockchain}
Lurdes~D Patr{\'\i}cio and Jo{\~a}o~J Ferreira.
\newblock Blockchain security research: theorizing through
  bibliographic-coupling analysis.
\newblock {\em Journal of Advances in Management Research}, 2020.

\bibitem{yli2016current}
Jesse Yli-Huumo, Deokyoon Ko, Sujin Choi, Sooyong Park, and Kari Smolander.
\newblock Where is current research on blockchain technology?—a systematic
  review.
\newblock {\em PloS one}, 11(10):e0163477, 2016.

\bibitem{huckle2016internet}
Steve Huckle, Rituparna Bhattacharya, Martin White, and Natalia Beloff.
\newblock Internet of things, blockchain and shared economy applications.
\newblock {\em Procedia computer science}, 98:461--466, 2016.

\bibitem{kosba2016hawk}
Ahmed Kosba, Andrew Miller, Elaine Shi, Zikai Wen, and Charalampos Papamanthou.
\newblock Hawk: The blockchain model of cryptography and privacy-preserving
  smart contracts.
\newblock In {\em 2016 IEEE symposium on security and privacy (SP)}, pages
  839--858. IEEE, 2016.

\bibitem{meiklejohn2013fistful}
Sarah Meiklejohn, Marjori Pomarole, Grant Jordan, Kirill Levchenko, Damon
  McCoy, Geoffrey~M Voelker, and Stefan Savage.
\newblock A fistful of bitcoins: characterizing payments among men with no
  names.
\newblock In {\em Proceedings of the 2013 conference on Internet measurement
  conference}, pages 127--140, 2013.

\bibitem{goldfeder2018cookie}
Steven Goldfeder, Harry Kalodner, Dillon Reisman, and Arvind Narayanan.
\newblock When the cookie meets the blockchain: Privacy risks of web payments
  via cryptocurrencies.
\newblock {\em Proceedings on Privacy Enhancing Technologies},
  2018(4):179--199, 2018.

\bibitem{siegel2016understanding}
David Siegel.
\newblock Understanding the dao attack.
\newblock {\em Retrieved June}, 13:2018, 2016.

\bibitem{bartoletti2020dissecting}
Massimo Bartoletti, Salvatore Carta, Tiziana Cimoli, and Roberto Saia.
\newblock Dissecting ponzi schemes on ethereum: identification, analysis, and
  impact.
\newblock {\em Future Generation Computer Systems}, 102:259--277, 2020.

\bibitem{nikolic2018finding}
Ivica Nikoli{\'c}, Aashish Kolluri, Ilya Sergey, Prateek Saxena, and Aquinas
  Hobor.
\newblock Finding the greedy, prodigal, and suicidal contracts at scale.
\newblock In {\em Proceedings of the 34th Annual Computer Security Applications
  Conference}, pages 653--663, 2018.

\bibitem{koteska2017blockchain}
Bojana Koteska, Elena Karafiloski, and Anastas Mishev.
\newblock Blockchain implementation quality challenges: a literature.
\newblock In {\em SQAMIA 2017: 6th Workshop of Software Quality, Analysis,
  Monitoring, Improvement, and Applications}, pages 11--13, 2017.

\bibitem{cai2017hardening}
Chengjun Cai, Xingliang Yuan, and Cong Wang.
\newblock Hardening distributed and encrypted keyword search via blockchain.
\newblock In {\em 2017 IEEE Symposium on Privacy-Aware Computing (PAC)}, pages
  119--128. IEEE, 2017.

\bibitem{beck2016blockchain}
Roman Beck, Jacob Stenum~Czepluch, Nikolaj Lollike, and Simon Malone.
\newblock Blockchain--the gateway to trust-free cryptographic transactions.
\newblock {\em ECIS 2016 Proceedings at AIS Electronic Library (AISeL)}, 2016.

\bibitem{herrera2016privacy}
Jordi Herrera-Joancomart{\'\i} and Cristina P{\'e}rez-Sol{\`a}.
\newblock Privacy in bitcoin transactions: new challenges from blockchain
  scalability solutions.
\newblock In {\em International Conference on Modeling Decisions for Artificial
  Intelligence}, pages 26--44. Springer, 2016.

\bibitem{reid2013analysis}
Fergal Reid and Martin Harrigan.
\newblock An analysis of anonymity in the bitcoin system.
\newblock In {\em Security and privacy in social networks}, pages 197--223.
  Springer, 2013.

\bibitem{koshy2014analysis}
Philip Koshy, Diana Koshy, and Patrick McDaniel.
\newblock An analysis of anonymity in bitcoin using p2p network traffic.
\newblock In {\em International Conference on Financial Cryptography and Data
  Security}, pages 469--485. Springer, 2014.

\bibitem{garay2015bitcoin}
Juan Garay, Aggelos Kiayias, and Nikos Leonardos.
\newblock The bitcoin backbone protocol: Analysis and applications.
\newblock In {\em Annual international conference on the theory and
  applications of cryptographic techniques}, pages 281--310. Springer, 2015.

\bibitem{lim2014analysis}
Il-Kwon Lim, Young-Hyuk Kim, Jae-Gwang Lee, Jae-Pil Lee, Hyun Nam-Gung, and
  Jae-Kwang Lee.
\newblock The analysis and countermeasures on security breach of bitcoin.
\newblock In {\em International conference on computational science and its
  applications}, pages 720--732. Springer, 2014.

\bibitem{kiktenko2018quantum}
Evgeniy~O Kiktenko, Nikolay~O Pozhar, Maxim~N Anufriev, Anton~S Trushechkin,
  Ruslan~R Yunusov, Yuri~V Kurochkin, AI~Lvovsky, and AK~Fedorov.
\newblock Quantum-secured blockchain.
\newblock {\em Quantum Science and Technology}, 3(3):035004, 2018.

\bibitem{rajan2019quantum}
Del Rajan and Matt Visser.
\newblock Quantum blockchain using entanglement in time.
\newblock {\em Quantum Reports}, 1(1):3--11, 2019.

\end{thebibliography}
\vspace{-.8cm}
\end{document}